\documentclass[aps,prl,11pt,amsmath,amssymb,reprint,nobibnotes,showkeys,superscriptaddress]{revtex4-1} 
\usepackage[utf8]{inputenc}
\usepackage[T1]{fontenc}

\usepackage[
    colorlinks=true,
    linkcolor=blue,
    citecolor=blue,
    urlcolor=blue
]{hyperref}

\newcommand{\dd}{\mathrm{d}}
\newcommand{\DD}{\mathrm{D}}

\begin{document}

\preprint{}

\title{The Lanczos potential in the Plebanski paradigm of gravity}

\author{Daniel Esp\'in}
 \email{danielesping@correo.ugr.es}
 \affiliation{Universidad de Granada, Granada-18071, Spain}
\author{Alejandro Jim\'enez Cano}
 \email{alejimcano@gmail.com}
 \affiliation{Escuela T\'ecnica Superior de Ingenier\'ia de Montes, Forestal y del Medio Natural, Universidad Politécnica de Madrid, 28040 Madrid, Spain}
\author{Javier Olmedo}
 \email{javolmedo@ugr.es}
 \affiliation{Departamento de F\'isica Te\'orica y del Cosmos, Universidad de Granada, Granada-18071, Spain}
\author{Ant\'onio Torres Manso}
 \email{antonio.manso@ijs.si}
 \affiliation{Jožef Stefan Institute, Jamova 39, 1000 Ljubljana,  Slovenia}

\begin{abstract}
    We present a new Plebanski formulation of General Relativity. The key difference is that we work with a gauge vector field derived from the Lanczos potential of the Weyl tensor. We show that, on shell, the gauge vector field encodes all nontrivial gravitational degrees of freedom and, in a suitable Lorentz gauge, Bianchi identities ensure that this gauge vector field satisfies a source-free wave equation for Ricci-flat spacetimes. This description suggests new ways of interpreting the true geometrical degrees of freedom with applications in both classical and quantum gravity.
\end{abstract}

\maketitle

\paragraph*{\bf{Introduction}}

The most common description of General Relativity theories assumes a Lorentzian metric on a 4D spacetime manifold along with a suitable connection. Other alternatives are based on tetrads and spin connections, as is the case of Ashtekar formulation of General Relativity \cite{Ashtekar:1986yd}. However, there are interesting formulations based on the one provided by Plebanski \cite{Plebanski:1977zz}, where the fundamental fields are a set of 2-forms, a spin connection, and a collection of scalar fields. The modern formulation of these theories belongs to the family of $BF$ theories \cite{Krasnov:2010olp, Freidel:2012np, Celada:2016jdt}. This framework is particularly useful because it actually allows one to cast Yang-Mills theories into a $BF$ description \cite{Martellini:1996ek, Escalante:2012wyy}. Indeed, it provides a first order formulation of this type of theories. In the 3D case, gravity turns out to be a pure topological $BF$ theory, with no dynamical degrees of freedom. Its quantization was discussed by Witten \cite{Witten:1988hc}. Nevertheless, gravity in 4D has local dynamical degrees of freedom that make the theory non-topological and highly nontrivial \cite{Capovilla:1989ac, Capovilla:1991kx, Capovilla:1991qb}.

Interestingly, a $BF$ formulation of classical General Relativity has served as starting point of a nonperturbative path integral quantization of gravity: the so-called spin foam approach \cite{Baez1999, Perez:2012wv, Livine2024}. Here, integrating out the collection of scalar fields (which are merely regarded as Lagrange multipliers) ensures that the $B$ fields satisfy the simplicity constraints, thereby allowing the introduction of a spacetime metric. In the classical theory, these scalars turn out to compose the pure gravitational sector, namely, the Weyl tensor. If they vanish in Ricci flat spacetimes, the curvature of the spin connection is flat. Therefore, treating these scalars as mere Lagrange multipliers seems awkward. However, this is not problematic in the quantum theory; integrating out these scalars is equivalent to imposing the simplicity constraints on the action, which reduces at the classical level to the Einstein-Hilbert action in terms of tetrads and spin connections.

In this letter we will show that these scalar fields can be replaced by covariant derivatives of a gauge vector potential with respect to the spin connection. We will discuss the equations of motion of the system, which correspond to i) compatibility of the spin connection with 2-forms, ii) the curvature of the connection being sourced by the geometrical (Lanczos gauge potential) and matter degrees of freedom, and iii) the equations of motion of gravitational degrees of freedom that result from Bianchi identities. Throughout this work we will assume Ricci flat spacetimes, namely, no presence of matter or cosmological constant. In this case, and in a suitable Lorentz gauge, the Lanczos gauge potential satisfies a simple wave equation with respect to the spin connection. Finally, we will comment on the advantages and shortcomings of this description for the classical and the quantum theory.  

~

\paragraph*{\bf{The Lanczos gauge potential\label{sec:ci}}}

Let us start introducing a collection of three complex 2-forms $\alpha^{I}{}_{ab}$ and the complex conjugated $\bar{\alpha}^{\dot{I}}{}_{ab}$, where the indices $I,J,...$  and $\dot{I},\dot{J},...$ run from 1 to 3 and are lowered and raised with the metrics $\delta_{IJ} = \mathrm{diag}(1,1,1)$ and $\delta_{\dot{I} \dot{J}} = \mathrm{diag}(1,1,1)$, respectively. We assume the internal gauge group to be $SU(2)$. On the other hand, the indices $a,b,\ldots$ correspond to spacetime abstract indices that run from 0 to 3. At this moment, it is convenient to assume that these 2-forms satisfy the usual simplicity constraints. Hence, they define a spacetime metric via the Urbantke expression
\begin{align}\label{eq:urban}
    \sqrt{|g|} g_{ab}=\frac{1}{12} \epsilon_{I J K} \epsilon^{c d e f} \alpha^I{}_{a c} \alpha^J{}_{d e} \alpha^K{}_{f b},
\end{align}
where $\epsilon$ denotes the totally antisymmetric Levi-Civita symbols, to be distinguished from the Levi-Civita tensors $\varepsilon$ used elsewhere. Equivalently, simplicity constraints guarantee that there is a basis of tetrads and the 2-forms can be decomposed as combinations of their wedge products \cite{Capovilla:1991kx}. Besides, these 2-forms are self-dual in the sense $({}^\star\alpha^I{}_{ab}\equiv)\frac{1}{2} \alpha^I{}_{cd} \varepsilon^{cd}{}_{ab} = -i \alpha^I{}_{ab}$, while $\bar{\alpha}^{\dot{I}}{}_{ab}$ turn out to be anti-self-dual. Moreover, the spacetime metric in Eq.~\eqref{eq:urban} can be used to lower and raise spacetime indices. With this, the following identity is satisfied:
\begin{equation}
    \alpha^I{}_{ac} \alpha^{Jc}{}_{b} = \delta^{IJ} g_{ab} +  {}^+\Sigma^{IJ}{}_{ab},
\end{equation}
with ${}^+\Sigma^{IJ}{}_{ab}\equiv  i \varepsilon^{IJK} \alpha_{Kab}$ the generators in the (1,0) spin representation of the Lorentz group. By complex conjugation, one arrives at similar identities for $\bar{\alpha}^{\dot{I}}{}_{ab}$, which are naturally associated to the generators in the (0,1) spin representation of the Lorentz group. Another interesting relation is 
\begin{align}\label{eq:projP}
    \alpha_{Iab} \alpha^{Icd} &= -4P^+_{ab}{}^{cd} = -2 \left(\delta^{[c}_{a}\delta^{d]}_{b} \pm \frac{i}{2}\varepsilon_{ab}{}^{cd}\right).
\end{align}
Hence, $P^+_{ab}{}^{cd}$ is the projector in the self-dual sector. Complex conjugation of Eq.~\eqref{eq:projP} gives the projector in the anti-self-dual sector. In addition, the self-dual and anti-self-dual sectors are orthogonal in the sense $\alpha_{Iab} \bar\alpha_{\dot{J}}{}^{ab} = 0$.

The Riemann tensor is completely determined by the Weyl curvature, i.e., $R_{abcd}=C_{abcd}$. The latter shares all symmetries of the Riemann tensor (antisymmetric in the first and second pairs of indices and symmetric in the exchange of the first and second pairs) and, in addition, is traceless, namely $C^c{}_{acb}=0$, even in non-Ricci flat spacetimes. Hence, it admits a natural decomposition $C_{abcd}=C^+_{abcd}+C^-_{abcd}$ where
\begin{align}\label{eq:Cp2H}
    C^+_{abcd} = H_{IJ}\alpha^I{}_{ab}\alpha^J{}_{cd},
\end{align}
is the self-dual part of the Weyl tensor, where $H_{IJ}$ is a symmetric and traceless $3\times3$ matrix of complex components. A similar expression can be found for the anti-self-dual component $C^-_{abcd}(=\overline{C^+_{abcd}})$ in terms of the complex conjugates $\bar{H}_{\dot{I}\dot{J}}$ and  $\bar\alpha^{\dot{I}}{}_{ab}$. Indeed, $H_{IJ}$ are linear combinations of Weyl scalars in the Newman-Penrose formalism~\cite{Newman1961}. Moreover, if we now consider the connection compatible with the 2-forms, namely ${D_{a} \alpha^I{}_{bc}=0}$ \footnote{
    The connection $\omega^I{}_a$ is determined by the condition $D_{a} \alpha_{Ibc}= \nabla_{a} \alpha_{Ibc} + i\varepsilon_{IJK}\omega^J{}_a\alpha^K{}_{bc}=0$, with $\nabla_{a}$ the usual Levi-Civita connection (acting only on spacetime indices).
    }, 
the Bianchi identity $D^aC^+_{abcd}=0$ now reads $\alpha^I{}_{ab} D^a H_{IJ} = 0$ \footnote{
    These equations are equivalent to those in Ref.~\cite{Newman1961}, concretely, see Eqs.~(4.5) in that reference. See also the analogy with electromagnetism in Eq.~(3.24) of Ref.~\cite{Agullo:2018nfv}.
    }.
From now on, we will use the compact notation $D_a T \equiv T_{;a}$ for any tensor $T$ with arbitrary spacetime and internal indices.

As we already mentioned above, the Weyl tensor in 4D admits a (real tensor) potential $L_{abc}$ originally introduced by Lanczos in Ref.~\cite{Lanczos1962} (see also Ref.~\cite{Dolan:1994yup} for more details). It satisfies the conditions
\begin{align}\label{eq:L-conds}
    L_{(ab)c} &= 0\,, \quad L_{[abc]}= 0, \, \quad L_{ac}{}^c=0.
\end{align}
One can easily verify that the Weyl tensor can be written as
\begin{align}
    C_{abcd}  = L_{ab[c;d]} + L_{cd[a;b]} - {}^{\star}L^{\star}_{ab[c;d]} - {}^{\star}L^{\star}_{cd[a;b]},\label{eq:WdecompL}
\end{align}
where, in the last two terms, the dual operation is applied separately to each pair of indices. Let us briefly comment that the Lanczos potential $L_{abc}$ satisfying conditions \eqref{eq:L-conds} have 16 independent components against 10 free components of the Weyl tensor. In order to (partially)
eliminate the extra freedom \footnote{
  In the literature it is common to adopt the Lorentz gauge in Eq.~\eqref{eq:LanczosDiv0} as a way to reduce the number of independent components from 16 to 10. However, this statement seems to miss a rigorous proof. See Ref.~\cite{Vishwakarma:2020yvo} for comments.
}, 
it is common to impose the Lorentz gauge
\begin{equation}\label{eq:LanczosDiv0}
    L_{abc}{}^{;c} = 0 .
\end{equation}
Since the Lanczos potential is antisymmetric in the first two indices, it naturally admits the following decomposition
\begin{align}\label{eq:labc-2-lia}
    L_{abc}=\alpha_{Iab}L^I{}_{c} + \bar{\alpha}_{{\dot{I}}ab} \bar{L}^{\dot{I}}{}_{c} \,,
\end{align}
with $L^I{}_{c}$ the Lanczos (complex) gauge potential that results after projecting the Lanczos potential on the basis of 2-forms. Concretely, $L^I{}_{a} = - \frac{1}{4} \alpha^{Ibc}L_{bca}$, and similarly for $\bar{L}^{\dot{I}}{}_{a}$. Since $L_{abc}$ is real, we have that $\bar{L}^{\dot{I}}{}_c = \overline{L^I{}_{c}}$.

Now, the second and third conditions in Eq.~\eqref{eq:L-conds} lead to (see \footnote{
    Notice that the cyclic condition $L_{[abc]} =0$ is equivalent to ${0={}^\star L_{ab}{}^b = (-i) (\alpha^I{}_{a}{}^{b} L_{Ib} - \bar{\alpha}^{\dot{I}}{}_{a}{}^{b} \bar{L}_{\dot{I}b})}$. This, together with the traceless condition ${0 = L^{ac}{}_c = \alpha^{Iac}L_{Ic} + \bar{\alpha}^{\dot{I}ac}\bar{L}_{\dot{I}c}}$ leads to Eq.~\eqref{eq:cyctrless}.
    })
\begin{equation} \label{eq:cyctrless}
    \alpha_I{}^{ac}L^I{}_{c} = 0, \quad  \bar{\alpha}_{\dot{I}}{}^{ac}\bar{L}^{\dot{I}}{}_{c} = 0,
\end{equation}
which are equivalent to
\begin{equation}
    L^I{}_a = i \varepsilon^{IJK} \alpha_{Ja}{}^b L_{Kb} ,\quad
    \bar{L}^{\dot{I}}{}_a = -i \varepsilon^{\dot{I}\dot{J}\dot K} \bar{\alpha}_{\dot{J}a}{}^b \bar{L}_{\dot Kb} \,.\label{eq:cyclicLIc}
\end{equation}
In terms of $L^I{}_a$, the Lorentz gauge in Eq.~\eqref{eq:LanczosDiv0} reads
\begin{equation}\label{eq:Lorenzgaugevec}
    L^I{}_a{}^{;a}=0,\quad \bar{L}^{\dot{I}}{}_a{}^{;a}=0.
\end{equation}

For the sake of brevity, we henceforth restrict our attention to the self-dual sector, since the corresponding expressions for the anti-self-dual sector are easily obtained by complex conjugation. We define the following self-dual 2-forms:
\begin{align}
    \mathcal{F}^I{}_{ab} 
        &\equiv \frac{1}{2} (L^I{}_{[a;b]} + i\, {}^{\star}L^I{}_{[a;b]})\,.
    \label{eq:Fiab-dual}
\end{align}
From Eq.~\eqref{eq:WdecompL}, it is not difficult to arrive at
\begin{align}
    C^+_{abcd}&= 2\alpha^{I}{}_{ab} \mathcal{F}_{Icd}+2\alpha^{I}{}_{cd} \mathcal{F}_{Iab}.
\end{align}
If we compare the previous expression with Eq.~\eqref{eq:Cp2H}, after applying suitable projections with the 2-forms, one finds:
\begin{equation}\label{eq:HcalF}
    H_{IJ} = - \alpha_{(I}{}^{ab} \mathcal{F}_{J)ab} \,.
\end{equation}
We must recall that $H_{IJ}$ is symmetric and traceless by construction \footnote{
    As a consistency test, one can check that  $ { H_{IJ}\delta^{IJ} = -\alpha_I{}^{ab} \mathcal{F}^I{}_{ab} = - \alpha_I{}^{ab} L^I{}_{a;b} = -(\alpha_I{}^{ab} L^I{}_{a})_{;b} = 0 \,,}$ by Eq.~\eqref{eq:cyclicLIc}.
    }.
Interestingly, one can check that $\varepsilon_{IJK} \alpha^{Iab} \mathcal{F}^{J}{}_{ab}$ is proportional to $L^I{}_a{}^{;a}$. Therefore, in the Lorentz gauge \eqref{eq:Lorenzgaugevec}, this quantity is vanishing, so $\alpha^{Iab} \mathcal{F}^{J}{}_{ab}$ becomes symmetric in $IJ$.

Let us now turn our attention to the Bianchi identity of the (self-dual component of the) Weyl tensor
\begin{align}
    0 = C_{abcd}^{+}{}^{;a}& = 2(\alpha^I{}_{ab} \mathcal{F}_{Icd}{}^{;a} + \alpha^I{}_{cd} \mathcal{F}_{Iab}{}^{;a}) \ .
\end{align}
If we project on $\alpha_J{}^{cd}$ the rightmost expression, we get 
\begin{align}\label{eq:bianchi-F}
    \alpha^J{}_{cd} \alpha_{Iab} \mathcal{F}^I{}_{cd}{}^{;a} - 4 \mathcal{F}^J{}_{ab}{}^{;a} = 0.
\end{align}
If we replace in terms of Lanczos gauge potentials, after some simple calculations, we get  
\begin{align}\label{eq:Dalambertiano-nogauge}
    2 L^I{}_{b;a}{}^{;a} -  i \varepsilon^{IJK} \alpha_{Jbc} L_{Ka}{}^{;a;c} - 2 L^I{}_a{}^{;a;b} = 0. 
\end{align}
Again, in the Lorentz gauge ($L^I{}_a{}^{;a}=0$), Eq.~\eqref{eq:bianchi-F} can be simplified to 
\begin{align}
    \mathcal{F}^J{}_{ab}{}^{;a} = 0,
\end{align}
and \eqref{eq:Dalambertiano-nogauge} can be written simply as $L^I{}_{b;a}{}^{;a}=0$. All the above arguments are equivalent for the anti-self-dual sector. In summary, we have
\begin{align}\label{eq:Dalambertiano}
  L^I{}_{b;a}{}^{;a}=0, \quad \bar{L}^{\dot{I}}{}_{b;a}{}^{;a}=0. 
\end{align}

At this point, there are several comments in order. The first one is that we have provided an alternative proof regarding the fact that the Lanczos potential satisfies the wave equation in Ricci flat spacetimes. Once we have demonstrated that both $L^{I}{}_{a}$ and $ \bar{L}^{\dot{I}}{}_{a}$ satisfy it, one can trivially check that $L_{bcd;a}{}^{;a}= 0$ by replacing Eqs.~\eqref{eq:labc-2-lia} and \eqref{eq:Dalambertiano}. If we look at the standard literature, one needs to prove that Eq.~(B15) in Ref.~\cite{Dolan:1994yup} is satisfied. There, a simple demonstration using spinorial formulation can be found. For an alternative proof using the standard formulation with spacetime indices, see Ref.~\cite{BrianEdgar1994}. The second comment is related to the fact that, as the Lanczos potential acts as a gauge vector potential, opens the possibility of writing gravitational tidal forces as line integrals of this gauge potential. This will be explored in future works. The third important comment is the relation between the matrix $H_{IJ}$ and Lanczos gauge potential in Eq.~\eqref{eq:HcalF}. To the best of our knowledge this relation has never been noticed in the literature of $BF$ theories. In this letter, we will show that replacing the scalars accompanying the simplicity constraints by (derivatives of) the Lanczos gauge potential provides an interesting new perspective in our understanding of gravity. 

~

\paragraph*{\bf{Lanczos $BF$ formulation}}

Let us consider the following $BF$ theory with the following fundamental complex variables: a $SU(2)$ gauge connection $A^I{}_a$, a 1-form $L^I{}_a$ and three 2-forms $\alpha^I{}_{ab}$ satisfying the simplicity constraints, i.e., they arise from a non-degenerate tetrad $\theta^{\rm A}$ (here ${\rm A}, {\rm B}...$ are frame indices in 4D). We introduce the derivative and curvature associated to $A^I{}_a$:
\begin{align}
    X^I{}_{a;b} &\equiv \nabla_{b}X^I{}_{a} + i \varepsilon^I{}_{JK} A^J{}_{b} X^K{}_{a},\\
    F^I{}_{ab}(A) &\equiv \partial_a A^I{}_b - \partial_b A^I{}_a+i \varepsilon^I{}_{JK} A^J{}_a A^K{}_b.
\end{align}
At this point, we will work with equations of motion. We leave the discussion of the action including the simplicity constraints to the end of this letter. Concretely, we have the following set of equations for the self-dual sector \cite{Freidel:2012np,Celada:2016jdt}
\begin{align}\nonumber
    &\alpha^I{}_{[bc;a]} = 0,\\\nonumber
    &F_{Iab}(A) + 2\Upsilon_{IJ} \alpha^J{}_{ab} = 0,\\
    & L^I{}_a = i \varepsilon^{IJK} \alpha_{Ja}{}^b L_{Kb} \, ,\label{eq:BF-EOM}
\end{align}
where we are using the shorthand 
\begin{equation}
     \Upsilon^{IJ}\equiv -\alpha^{(I|ab}L^{|J)}{}_{[a;b]}. \label{eq:Psi}
\end{equation}
The first line in Eq.~\eqref{eq:BF-EOM} states that the 2-forms $\alpha^I{}_{ab}$ have vanishing exterior covariant derivative with respect to $A^I{}_a$. Together with the simplicity constraints, this condition identifies $A^I{}_a$ with the self-dual part of the torsion-free spin connection associated with the tetrad $\theta^{\rm A}$. Accordingly, $F^I{}_{ab}$ corresponds to $R_{abcd} \alpha^{Icd}/2$. The third line in Eq.~\eqref{eq:BF-EOM} imposes the appropriate conditions on the gauge vector potential $L^I{}_a$, ensuring that the Weyl tensor constructed from it has the required algebraic properties. Finally, the second line in Eq.~\eqref{eq:BF-EOM} can be expressed on shell as 
\begin{align}\label{eq:FtocalF}
    F^I{}_{ab}(A) = \alpha^{J}{}_{ab} (\alpha^{Icd}\mathcal{F}_{Jcd}+\alpha_{J}^{cd} \mathcal{F}_{Icd}).
\end{align}
In the Lorentz gauge (see comment after Eq.~\eqref{eq:HcalF}),
this amounts to
\begin{align}\label{eq:FtoDL}
    F^I{}_{ab}(A)= - 8 \mathcal{F}^I{}_{ab} =- 8 P^+_{ab}{}^{cd} L^I{}_{[c;d]},
\end{align}
where we have used the projector on the self-dual sector defined in Eq.~\eqref{eq:projP}. In essence, Eq.~\eqref{eq:FtoDL} suggests a simple picture in which the gravitational degrees of freedom encoded in the gauge vector potential $L^I{}_a$ source the spacetime curvature encoded in $F^I{}_{ab}(A)$. In particular, setting $L^I{}_a=0$ leads to a vanishing curvature $F^I{}_{ab}(A)$ (i.e.,  a flat connection $A^I{}_a$) and, after imposing the corresponding reality conditions, to a locally flat spacetime metric. Conversely, a nontrivial $L^I{}_a$ can give rise to a nonvanishing curvature and hence to a nontrivial geometry. This is the most striking point of this formulation!

The previous conclusions allow us to interpret the true gravitational degrees of freedom to be codified into a dynamical gauge vector potential $L^I{}_a$, which are explicitly separated from the spacetime geometry codified in the connection $A^I{}_a$ and the three 2-forms $\alpha^I{}_{ab}$. Of course, they are dynamically coupled by expressions in Eq.~\eqref{eq:BF-EOM}, in the same way that the usual Einstein equations dynamically couple geometry and matter. The Bianchi identity for the curvature of $A^I{}_a$ reads $D_{[a}F^I{}_{bc]}=0$ and, since $F^I{}_{ab}$ belongs to the self-dual sector, this identity is equivalently expressed as $F^I{}_{ab}{}^{;b}=0$. This therefore implies Eq.~\eqref{eq:Dalambertiano-nogauge} for the gauge vector potential $L^I{}_a$. Besides, these equations of motion remain invariant if we choose the Lorentz gauge in Eq.~\eqref{eq:Lorenzgaugevec}, hence we recover Eq.~\eqref{eq:Dalambertiano} for the self-dual (Ricci flat) sector.

This description and the Einstein theory of relativity show the same parallelism: Einstein equations couple gravity and matter, but one should also account for the equations of motion of the matter content via conservation of stress-energy tensor. Here, the Bianchi identities provide the ``conservation law'' for the Weyl tensor and, consequently, for its Lanczos potential. Let us note that it satisfies a wave equation with respect to the connection $A^I{}_a$. Therefore, it seems well-defined and, in principle, solvable (likely numerically) provided suitable initial data for $L^I{}_a$ and its derivatives. Moreover, one must also solve Eq.~\eqref{eq:FtoDL}. However, since the initial data is codified in the Lanczos potential, Eq.~\eqref{eq:FtoDL} serves to compute $A^I{}_a$ out of $L^I{}_a$. No additional initial information seems to be required. In summary, the most important take home message is that in 4D gravity, the gravitational degrees of freedom can be codified into a dynamical gauge (Lanczos) potential sourcing the spacetime geometry via Eqs.~\eqref{eq:BF-EOM}.

~

\paragraph*{ \bf{Lanczos $BF$ Action}}

The final question we want to answer in this letter is whether there is an action from which we can derive the above set of equations of motion. The answer is in the affirmative. We want a Lagrangian similar to the one in Eq.~(6) of Ref.~\cite{Celada:2016jdt}. 

To that end, we will consider an action of the form:
\begin{equation}
    S [B, \theta, A, L, \lambda, \sigma, \chi]= \int_{\mathcal{M}} \mathbb{L} ,
\end{equation}
where the Lagrangian 4-form is given by
\begin{align}\nonumber \label{eq:BFaction}
    \mathbb{L} 
    &= B^{I}\wedge F_{I}(A) + C_{IJ}(B)\, \Upsilon^{IJ}(\theta, A, L) + \lambda_I \wedge  \DD  \alpha^I(\theta) \\
    &\quad + \chi_I \wedge \Big(B^I - \alpha^I(\theta)\Big) + \sigma_I \wedge N^I(L,\theta) + \textrm{c.c.},
\end{align}
with the abbreviations:
\begin{align}
    F^I(A) &\equiv \dd A^I +\frac{i}{2} \varepsilon^I{}_{JK} A^J \wedge A^K ,\\
    C_{IJ}(B) &\equiv B_{I}\wedge B_{J}-\frac{1}{3} \delta_{IJ}B_{K}\wedge B^{K} ,\\
    \Upsilon^{IJ}(\theta,A,L) &\equiv - {}^\star \Big(\alpha^{(I}\wedge {}^\star \DD L^{J)}\Big) \label{eq:Psi_difform},\\
    N^I(L,\theta) &\equiv L^I - i \varepsilon^{IJK} \, {}^\star \big(L_K \wedge {}^\star \alpha_J\big) .
\end{align}
A few remarks are in order at this point:
\begin{itemize}

    \item The star correspond to the Hodge dual with respect to the volume of the coframe (tetrad) $\theta^{\rm A}$ or, equivalently, to the spacetime metric $g_{ab} = \eta_{{\rm AB}} \theta^{\rm A}{}_a \theta^{\rm B}{}_b$.
    
    \item We work, for convenience, in the language of differential forms. The ranks of the fundamental variables are summarized in Table~\ref{tab:rangos-formas}.
    \renewcommand{\arraystretch}{1.5}
    \begin{table}
    \centering
    \begin{tabular}{c|c}
        \hline
        $\theta^{\rm A}$, $A^I$, $L^I$, $\lambda_I$ & 1-forms \\
        \hline
        $B^I$, $\chi_I$ & 2-forms \\
        \hline
        $\sigma_I$ & 3-form \\
        \hline
    \end{tabular}
    \caption{Rank of the fundamental variables.}
    \label{tab:rangos-formas}
    \end{table}
    \renewcommand{\arraystretch}{1}

    \item The complex conjugate (anti-self-dual) sector is decoupled from the self-dual variables shown in \eqref{eq:BFaction}. In practical terms, it can be ignored when computing variations.

    \item $\DD$ is the exterior covariant derivative associated to $A^I$. For any 1-form $X$, this corresponds to $(\DD X)_{ab} = 2 X_{[b;a]}$ in tensor notation.

    \item The 2-forms $\alpha^I$ are the ones introduced at the beginning of this text and are derived from $\theta^{\rm A}$. They automatically satisfy the simplicity constraints \footnote{
        In particular, they have the form $\alpha^I( \theta) = \Pi^I{}_{\rm AB} \theta^{\rm A}\wedge \theta^{\rm B}$, where the projector $\Pi$ is independent of the field variables of the theory.
    }.
    
    \item The 4-form $C_{IJ}$ corresponds to the simplicity constraints of $B^I$.

    \item The 0-form $\Upsilon^{IJ}$ is exactly the one introduced in Eq.~\eqref{eq:Psi}.

    \item The 1-form $N^{I}$ has the following components:
    \begin{equation}
        N^I{}_a = L^I{}_a - i \varepsilon^{IJK} \alpha_{Ja}{}^b L_{Kb} , 
        \label{eq:Ncomp}
    \end{equation}
    namely, $N^I = 0$ encodes the condition \eqref{eq:cyclicLIc}.
\end{itemize}

In the following we will prove that the branch of solutions in which the three Lagrange multipliers ($\lambda_I$, $\chi_I$ and $\sigma_I$) are set to zero reproduces Eqs.~\eqref{eq:BF-EOM}.

First, we compute the equations of motion of the Lagrange multipliers $\lambda_I$, $\chi_I$ and $\sigma_I$, which are respectively given by:
\begin{align}
    \DD \alpha^I &= 0 \label{eq:DB}, \\
    B^I &= \alpha^I \label{eq:Balph},\\
    N^I & = 0 \label{eq:N}\,.
\end{align}
The first two imply that, on shell, we can always take
\begin{equation}
    \DD B^I = 0\,,\qquad C_{IJ}(B) = C_{IJ}(\alpha) \equiv 0\,,\label{eq:simp}
\end{equation}
(the latter is true by definition of $\alpha^I$). On the other hand, Eq.~\eqref{eq:N} enforces the cyclic and traceless condition Eq.~\eqref{eq:cyclicLIc}.

The equations of motion of $B^I$, $A^I$ and $L^I$ are, respectively:
\begin{align}
    0 & = F_I + 2\Upsilon_{IJ} \alpha^J + \chi_I , \nonumber \\
    0 & = \DD B_I - i({}^\star C_{KJ}) \varepsilon^J{}_{IL}\, L^L\wedge {}^\star \alpha^K  + i \varepsilon_{IJK} \lambda^J \wedge \alpha^K , \nonumber \\
    0 & = \DD \big(({}^\star C_{IJ})\ {}^\star \alpha^J \big) + \sigma_I - i \varepsilon_I{}^{JK}\ {}^\star \alpha_J \wedge{}^\star\sigma_K , 
\end{align}
After applying the simplifications \eqref{eq:simp} and  selecting the branch with vanishing Lagrange multipliers, the last two trivialize and the first one becomes:
\begin{align}
    0 & = F_I + 2\Upsilon_{IJ} \alpha^J\,.  \label{eq:EoMA}
\end{align}

Finally, regarding the equation of motion of the tetrad, it is not difficult to check that it consists of terms that are either proportional to $C_{IJ}$ or to one of the Lagrange multipliers, so this equation is automatically satisfied in the selected branch.

In summary, among the 7 equations of motion, 3 of them trivialize in the chosen branch, 1 leads to the on-shell identification of $B^I$ and $\alpha^I$, and the remaining 3 (namely, Eqs.~\eqref{eq:DB}, \eqref{eq:N} and \eqref{eq:EoMA}) reproduce the desired dynamics in Eqs.~\eqref{eq:BF-EOM}. 

Therefore, we have derived an action that reproduces the Plebanski formulation of General Relativity using a gauge vector field with suitable kinetic terms, rather than relying on mere spectator scalar fields. More importantly, gravitational (physical) degrees of freedom can be naturally codified in this gauge vector field rather than in the connection and 2-forms, both being sourced by the former via Eqs.~\eqref{eq:BF-EOM}.

~

\paragraph*{\bf{Discussion}}

In this letter we have presented a reformulation of vacuum General Relativity as a $BF$ theory in which the symmetric traceless matrix $\psi_{ij}$ of the standard Plebanski formulation \cite{Celada:2016jdt,Perez:2012wv}, usually treated as a set of Lagrange multipliers enforcing the simplicity constraints, is replaced by covariant derivatives of a gauge vector field $L^I{}_a$ derived from the Lanczos potential of the Weyl tensor. On shell, this Lanczos gauge potential encodes all nontrivial gravitational degrees of freedom: it sources the curvature of the connection $A^I$ via Eq.~\eqref{eq:BF-EOM}, while the Bianchi identities describe its dynamics, reducing in the Lorentz gauge to the source-free wave equation~\eqref{eq:Dalambertiano} for Ricci flat spacetimes. This mirrors the structure of the Einstein equations with matter: geometry (the connection $A^I$ and the 2-forms $\alpha^I$) is sourced by $L^I{}_a$, whose own evolution follows from a Bianchi identity that resembles a conservation law. We have also derived a Lagrangian, Eq.~\eqref{eq:BFaction}, from which this set of equations of motion follows, after a suitable restriction of the space of solutions. This result places the construction on the same variational footing as standard $BF$ formulations. Additionally, we obtained an alternative proof that the Lanczos potential obeys the wave equation in Ricci-flat spacetimes, together with the relation~\eqref{eq:HcalF} between $H_{IJ}$ and the Lanczos gauge potential, which to our knowledge had not been noticed in the $BF$ literature.

We conclude this manuscript by outlining some limitations and future research directions. Unlike classical 4D General Relativity, which possesses well-established uniqueness theorems in vacuum and for certain matter content, our formulation currently lacks explicit uniqueness results. However, the simplicity of the equations of motion makes proving such theorems a natural next step. In any case, our description reproduces General Relativity in vacuum. Adding matter, we can follow the same ideas of \cite{Tennie:2010qq,Hughes:2026lif}. Moreover, our results are valid only in 4D gravity, although there are generalizations of Lanczos potentials in larger dimensions \cite{Edgar:2004iq} that provide a natural future extension of our work. In addition, let us recall that General Relativity written as a $BF$ theory with simplicity constraints is the starting point of most of the spin foam models in quantum gravity \cite{Perez:2012wv}. They treat the components of the matrix $\psi_{ij}$ as Lagrange multipliers (or spectator fields) that are integrated out in the path integral, resulting in projectors of simplicity constraints on the Hilbert space of the theory. If one wants to follow the same strategy here with a Lanczos potential, one has to deal not only with the simplicity constraints but also the restrictions on the Lanczos potential. One of the advantages is that $\alpha^I{}_{ab}$ comes with extra structure: the tetrad from which they can be derived if the simplicity constraints are satisfied. In this sense, condition $\alpha^I =B^I$ guarantees that the simplicity constraints of $B^I$ are satisfied selecting the right topological/geometric sector \cite{Engle:2012yg}. Hence, one does not need to solve them explicitly, which is one of the difficulties of the spin foam program. Finally, our results might also find applications in classical and quantum Regge calculus~\cite{Williams1992}. 

~

{\it \textbf{Acknowledgments}:} We want to acknowledge A. del R\'io, D. Oriti, R. Fonseca, B. Janssen and A. F. Sanfilippo for fruitful discussions. Financial support is provided by the Spanish Government through the project PID2022-140831NB-I00 funded by MI-CIU/EI/10.13039/501100011033 and FEDER, UE.  ATM is supported by the Slovenian Quantum Science Hub co-funded by the Marie Skłodowska-Curie Actions programme (GA-101177446) and the Slovenian Research and Innovation Agency (ARIS), contract number 5110-18/2025-5.

\bibliography{references}

\onecolumngrid

\appendix
\section{Exterior algebra: useful expressions}

Take a general frame $e_{\rm A}$ with dual coframe $\theta^{\rm A}$ on an $n$-dimensional manifold $(\mathcal{M}, g)$ (with arbitrary signature). Let $X$ and $Y$ be arbitrary $p$-forms on this manifold, and let ${}^\star$ be the Hodge dual associated with the volume form of $g$, i.e., 
\begin{equation}
    {}^\star 1 = {\rm vol}_g \equiv \frac{1}{n!}  \varepsilon_{{\rm A}_1...{\rm A}_n} \theta^{{\rm A}_1}\wedge \ldots \wedge  \theta^{{\rm A}_{n}}.
\end{equation}
If we use the convention
\begin{equation}
    {}^\star X = \frac{1}{p!(n-p)!} X_{{\rm A}_1...{\rm A}_p} \varepsilon^{{\rm A}_1...{\rm A}_p}{}_{{\rm B}_1...{\rm B}_{n-p}} \theta^{{\rm B}_1}\wedge \ldots \wedge  \theta^{{\rm B}_{n-p}}
\end{equation}
the following identities hold:
\begin{align}
    e_{\rm A} \lrcorner \, {}^\star X &= {}^\star (X \wedge \theta_{\rm A}) ,\\
    \theta^{\rm A} \wedge {}^\star X &= (-1)^{p+1}\, {}^\star (e^{\rm A} \lrcorner X) ,\\
    {}^\star{}^\star X &= (-1)^{p(n-p)} {\rm sign}(\det(g)) X ,\\
    {}^\star {\rm vol}_g &= {\rm sign}(\det(g)) ,\\
    X\wedge {}^\star Y &= Y\wedge {}^\star X,\\
    {}^\star(X\wedge {}^\star Y) &= \frac{1}{p!} {\rm sign}(\det(g)) X^{{\rm A_1...A_p}} Y_{{\rm A_1...A_p}},
\end{align}
where $\lrcorner$ denotes the internal product. If we focus on $n = 4$ and assume that the metric is Lorentzian with signature $(-+++)$, we can prove:
\begin{equation}
    e_{\rm A} \lrcorner \, {}^\star(\theta^{\rm B} \wedge {}^\star Y) = Y_{\rm A}{}^{\rm B}\,.
\end{equation}

These identities have been used to construct the differential form $N^I$ (to reproduce Eq.~\eqref{eq:Ncomp}), the differential form expression of Eq.~\eqref{eq:Psi} given in Eq.~\eqref{eq:Psi_difform}, and the derivation of the equations of motion.

\end{document}